\documentstyle[pre,floats,twocolumn,aps]{revtex}

\begin{document}
\draft
\date{January 5, 1999}
\title{Absence of Chaos in Bohmian Dynamics}
\author{Sheldon Goldstein\thanks{Electronic address: oldstein@math.rutgers.edu}}
\address{Department of Mathematics, Rutgers University, Piscataway, NJ
08854-8019}
\maketitle
\begin{abstract}
The Bohm motion for a particle moving on the line in a quantum state
that is a superposition of $n+1$ energy eigenstates is quasiperiodic with
$n$ frequencies.
\end{abstract}
\pacs{PACS number(s): 05.45.+b, 03.65.Bz, 03.65.Ge}

In a recent paper \cite{bonfim}, O. F. de Alcantara Bonfim, J. Florencio,
and F. C. S\'a Barreto claim to have found numerical evidence of chaos in
the motion of a Bohmian quantum particle in a double square-well potential,
for a wave function that is a superposition of five energy eigenstates. But
according to the result proven here, chaos for this motion is
impossible. We prove in fact that for a particle on the line in a
superposition of $n+1$ energy eigenstates, the Bohm motion $x(t)$ is always
quasiperiodic, with (at most) $n$ frequencies. This means that there is a
function $F(y_1, \dots, y_n)$ of period 2$\pi$ in each of its variables and
$n$ frequencies $\omega_1, \dots, \omega_n$ such that $x(t)= F(\omega_1 t,
\dots, \omega_n t).$

The Bohm motion for a quantum particle of mass $m$ with wave function
$\psi=\psi(x,t)$, a solution to Schr\"odinger's equation, is defined by
\begin{equation}\label{be}
dx/dt= (\hbar/m) \mbox{Im}\nabla \psi/\psi.
\end{equation} 
The right hand side of (\ref{be}) depends upon $\psi$ only through its
associated ray. In particular, if the wave function 
\begin{equation}\label{sup}
\psi(x,t) = \Sigma_{i=0}^n a_ie^{-iE_i t/\hbar}\phi_i(x)
\end {equation}
is a superposition of $n+1$ energy eigenstates $\phi_i$, then the right
hand side of (\ref{be}) is, in its dependence upon $t$, quasiperiodic with
$n$ frequencies, as is $|\psi|$. 

The quasiperiodicity in time of the vector field defining a dynamical
system in general does not imply any corresponding property of the motion,
since an autonomous system (one defined by a time independent vector field)
can be chaotic. (In fact, it is autonomous systems that are normally
studied in chaos theory.) However, for the Bohm motion on the line, the
position of the particle is anchored in the (normalized) wave function, in
such a way that its motion $x(t)$ inherits the quasiperiodicity of
$|\psi|$:

A crucial feature of the motion (\ref{be}) is the equivariance of $|\psi|^2$,
i.e., the fact that probabilities for configurations given by
$|\psi(x,t)|^2$ are consistent with the dynamics (\ref{be}). This is  a
completely general feature of the  Bohmian dynamics, valid in any dimension
for any wave function satisfying Schr\"odinger's equation. For a single
particle moving on the line, it has the following important consequence:
\begin{equation}\label{nd}
\int_{-\infty}^{x(t)}|\psi(x',t)|^2\,dx'=\int_{-\infty}^{x(0)}|\psi(x',0)|^2\,dx',
\end{equation}
which follows from equivariance since in one-dimension the dynamics is
order-preserving, and in particular the evolution from time 0 to time $t$
carries the interval $(-\infty, x(0))$ to $(-\infty, x(t))$. 

Given $\psi(x,0)$ and $x(0)$, equation (\ref{nd}) determines $x(t)$ as a
functional of $|\psi(x,t)|^2$, and thus $x(t)$, like $|\psi(x,t)|^2$, is
quasiperiodic with $n$ frequencies. In fact $x(t)= F(\omega_1 t, \dots,
\omega_n t)$ with $\omega_i=(E_i-E_0)/\hbar$ for $i=1,\dots,n$ and $F(y_1,
\dots, y_n)=G(\int_{-\infty}^{x(0)}|\psi(x',0)|^2\,dx')$ where $G$ is the
inverse of the function $H(x)=\int_{-\infty}^{x}|\psi(x')|^2\,dx'$ with
\begin{equation}
\psi(x)\equiv\psi_{y_1,\dots,y_n}(x)= a_0\phi_0(x)+\Sigma_{i=1}^n a_ie^{-iy_i}\phi_i(x).
\end{equation} 
	
For the one-dimensional motion $x(t)$ the Lyapunov exponent $\lambda$ is
given by
\begin{equation}
\lambda=\lim_{t\to\infty}t^{-1}\ln {dx(t)\over dx(0)}.
\end{equation}
It presumably follows from the quasiperiodicity of $x(t)$ alone that
$dx(t)/dx(0)$ is similarly  quasiperiodic. In any case, we have by
equivariance that $|\psi(x(t),t)|^2dx(t)=|\psi(x(0),0)|^2dx(0)$, so that 
\begin{equation}\label{dx} 
 dx(t)/dx(0)={|\psi(x(0),0)|^2\over |\psi_{y_1,\dots,y_n}(F(y_1,\dots,y_n))|^2}
\end{equation} 
with $y_i=\omega_i t$. Hence  $dx(t)/dx(0)$ is quasiperiodic with $n$
frequencies and thus $\lambda=0$.

Remarks: (i) In one-dimension we always have that
$dx(t)/dx(0)=|\psi(x(0),0)|^2/|\psi(x(t),t)|^2$. Thus the vanishing of the
Lyapunov exponent $\lambda$ is more general than described here, and should
be valid for any wave function, on the circle as well as the line. After
all, for bound states the ratio on the right is not likely to grow or
decrease in any systematic way at all, while for states with continuous
spectrum the behavior will be at most power law; in no case will there be
exponential growth or decay. (ii) Another aspect of chaos, the weak
convergence of densities to the ``equilibrium'' distribution (for Bohmian
mechanics given by $|\psi(x(t),t)|^2$) will, as a simple consequence of the
order preserving character of such motions, almost always fail for any
one-dimensional flow, Bohmian or otherwise. The sole exception can occur
only when the asymptotic ``equilibrium'' distribution is concentrated on a
single (perhaps moving) point, something that is impossible for Bohmian
mechanics.

I am grateful to Michael Kiessling for helpful suggestions. This work was
supported in part by NSF Grant No.  DMS-9504556.

\end{document}